\documentclass[aps,prd,reprint,onecolumn, groupedaddress,longbibliography, 12pt]{revtex4-1}

\usepackage{amsmath,amssymb}
\usepackage{graphicx}
\usepackage{color}
\usepackage{caption}
\usepackage{hyperref}
\usepackage{enumerate}
\begin{document}

\title{Vorticity in analogue gravity}


\author{Bethan Cropp}
\email[]{bcropp@issertvm.ac.in}
\affiliation{SISSA, Via Bonomea 265, 34136 Trieste, Italy, \\
     INFN sezione di Trieste, Via Valerio 2, 34127 Trieste, Italy
                       and \\
School of Physics, Indian Institute of Science Education and Research Thiruvananthapuram (IISER-TVM), Trivandrum 695016, India.}
\author{Stefano Liberati}
\email[]{stefano.liberati@sissa.it}
\affiliation{SISSA, Via Bonomea 265, 34136 Trieste, Italy
                                       and \\
     INFN sezione di Trieste, Via Valerio 2, 34127 Trieste, Italy.}
\author{Rodrigo Turcati}
\email[]{rturcati@sissa.it}
\affiliation{SISSA, Via Bonomea 265, 34136 Trieste, Italy	
                                       and \\
     INFN sezione di Trieste, Via Valerio 2, 34127 Trieste, Italy.}

\def\d{{\mathrm{d}}}
\newcommand{\scri}{\mathscr{I}}
\newcommand{\sun}{\ensuremath{\odot}}
\def\J{{\mathscr{J}}}
\def\L{{\mathscr{L}}}
\def\sech{{\mathrm{sech}}}
\def\T{{\mathcal{T}}}
\def\tr{{\mathrm{tr}}}
\def\diag{{\mathrm{diag}}}
\def\ln{{\mathrm{ln}}}
\def\Horava{Ho\v{r}ava}
\def\Aether{\AE{}ther}
\def\AEther{\AE{}ther}
\def\aether{\ae{}ther}
\def\UH{{\text{\sc uh}}} 
\def\KH{{\text{\sc kh}}}


\begin{abstract}

In the analogue gravity framework, the acoustic disturbances in a moving fluid can be described by an equation of motion identical to a relativistic scalar massless field propagating in a curved spacetime. This description is possible only when the fluid under consideration is barotropic, inviscid and irrotational. In this case, the propagation of the perturbations is governed by an acoustic metric which depends algebrically on the local speed of sound, density and the background flow velocity, the latter assumed to be vorticity free. In this work we provide an straightforward extension in order to go beyond the irrotational constraint. Using a charged --- relativistic and non-relativistic --- Bose--Einstein condensate as a physical system, we show that in the low momentum limit and performing the eikonal approximation we can derive a d'Alembertian equation of motion for the charged phonons where the emergent acoustic metric depends on a flow velocity in the presence of vorticity.

\end{abstract}

\maketitle

\tableofcontents

\section{Introduction}

Since the seminal work of Unruh \cite{Unruh:1980cg}, and its subsequent development \cite{Visser:1993ub,Visser:1997ux}, the analogue gravity framework has become a field of intense investigation over the last years. Issues such as Hawking radiation \cite{Unruh:1980cg,Barcelo:2003wu}, cosmological particle production \cite{Weinfurtner:2008ns,Weinfurtner:2008if}, emergence of spacetime and the fate of Lorentz invariance at short distances have acquired new insights within the analogue gravity field (for a general review, see \cite{Barcelo:2005fc}). 

Analogue gravity was born out of the realization that acoustic disturbances in flowing fluids can, under suitable conditions, be described in the formalism of curved spacetime. In fact, when taking into account linearized perturbations over barotropic irrotational moving fluids, it can be demonstrated that the phonon excitations propagate under an effective acoustic metric, which is fully characterized by the background quantities of the flow. This remarkable result is valid both in relativistic and non-relativistic cases \cite{Unruh:1980cg,Visser:1997ux,Visser:2010xv,Moncrief:1980,Bilic:1999sq}.  

Among the several condensed-matter and optical systems considered in analogue gravity, Bose--Einstein condensates (BECs) \cite{Barcelo:2003wu,Barcelo:2000tg,Lahav:2009wx,Jannes:2009yr,Mayoral:2010ck} are of particular interest as they provide simple low temperature systems with a high level of quantum coherence. The phonons in the BECs are treated as quantum perturbations over a classical background condensate, so providing an attractive scenario to describe semiclassical gravity phenomena, such as acoustic Hawking radiation \cite{Unruh:1980cg,Garay:2000jj,Garay:1999sk}. BECs also have inspired many interesting analogue models for the emergence of symmetries, spacetime and more recently a mechanism of emergent dynamics \cite{Belenchia:2014hga}.

It is worth noticing that all these analogue systems leading to these striking results, are characterized by the requirement of zero vorticity, i.e., in their thermodynamic limit they all provide irrotational fluids, 
for instance, in section 2.3 of \cite{Barcelo:2005fc}. In the case of BECs this is automatically imposed by the fact that the associated current is the derivative of a scalar, i.e., the phase of wavefunction of the condensate. Some works have tried to introduce acoustic metrics for different classes of fluid flows. An important contribution in this subject was made in \cite{PerezBergliaffa:2001nd} and we shall discuss the relative difference between our and their finds later on. 

Our aim in this paper is demonstrate that we can, in a charged condensate, for sufficiently low momenta, find a wave equation to describe the propagation of phonons under an effective acoustic metric in the presence of vorticity. First, we investigate the behavior of the mode excitations in a charged condensate. Analyzing the charged fluid we conclude that the electromagnetic minimal coupling introduces vorticity in the system. In the appropriate limit, we demonstrate that it is possible to build an acoustic metric which takes into account vorticity. 
We perform this analysis in the relativistic and non-relativistic cases.

The paper is arranged as follows: In section II we describe how the condensation of a charged relativistic gas can be formed. In section III, we analyze the perturbations over the top of the charged condensate. The relativistic acoustic metric in the presence of vorticity is derived in section IV. In section V, we show that the same results appear in the non-relativistic condensate. Our conclusions and remarks are present in section VI.

In our conventions the signature of the metric is (-,+,+,+).


\section{Charged relativistic Bose--Einstein condensate}\label{coupledrbec}

Let us start by considering the general theory of Bose--Einstein condensation of a charged relativistic ideal gas (for a general review, see \cite{Kapusta:1981aa}). The $U(1)$ gauge invariant Lagrangian describing the interaction of the complex scalar field ${\phi}$ and the electromagnetic field $A^{\mu}$ can be written as
\begin{eqnarray}\label{lagrangiandensity}
\mathcal{L}=-\eta^{\mu\nu}\left(D_{\mu} {\phi}\right)^{*}\left(D_{\nu} {\phi}\right)-\frac{m^{2}c^{2}}{\hbar^{2}} {\phi}^{*}{\phi}-\lambda( {\phi}^{*} {\phi})^{2}-\frac{1}{4}F_{\mu\nu}F^{\mu\nu},
\end{eqnarray}
where $F_{\mu\nu}(\equiv\partial_{\mu}A_{\nu}-\partial_{\nu}A_{\mu})$ is the electromagnetic field strength tensor, $m$ is the mass of the bosons, $\lambda$ is a coupling constant and the covariant derivative is defined by $D_{\mu}=\partial_{\mu}+\frac{iq}{c\hbar}A_{\mu}$. The Noether theorem leads to a locally conserved current $j^{\mu}$ which is given by
\begin{equation}
 {j^{\mu}}=\frac{i}{2}\left[ {\phi}\left({D}^{\mu}{\phi}\right)^{*}- {\phi}^{*}\left(D^{\mu}{\phi}\right)\right]. 
\end{equation}
The conserved charge associated to local $U(1)$ symmetry is
\begin{equation}
\mathcal{Q}=i\int{d^{3}x}\left[ {\phi}({D}_{0} {\phi})^{*}- {\phi}^{*}\left({D}_{0} {\phi}\right)\right], 
\end{equation}
which has an associated bosonic chemical potential, $\mu$. The momentum canonically conjugate to ${\phi}$ is
\begin{equation}
 {\pi}=\frac{\partial\mathcal{L}}{\partial\dot{ {\phi}}}=\dot{ {\phi}}^{*}+i\frac{q}{\hbar{c}}A_{0} {\phi}^{*}. 
\end{equation}

Since we are treating the fields $ {\phi}$ and $ {\phi}^{*}$ independently, the Hamiltonian density will be given by
\begin{equation}
\mathcal{H}={\pi}\dot{ {\phi}}+{\pi}^{*}\dot{ {\phi}}^{*}-\mathcal{L}. 
\end{equation}

The partition function is defined by
\begin{eqnarray}
\mathcal{Z}&=&\mathcal{N}\int\left(DA\right)\left({D}{\pi}\right)^{*}\left({D}{\pi}\right)\left({D}{\phi}\right)^{*}\left({D}{\phi}\right)\mathrm{exp}\left\lbrace\int^{\beta}_{0}d\tau\int{d^{3}x}\left[{\pi}\dot{{\phi}}+{\pi}^{*}\dot{{\phi}}^{*}-\left(\mathcal{H}-\mu{Q}\right)\right]\right\rbrace\times\nonumber\\
&&\times{\mathrm{det}\left(\frac{\partial{F}}{\partial{\omega}}\right)\delta{F}}. 
\end{eqnarray}

Integrating the momenta away, we arrive at
\begin{eqnarray}\label{partitionfunction}
\mathcal{Z}=\mathcal{N}\int\left(DA\right)\left(D{\phi}\right)^{*}\left(D{\phi}\right)\mathrm{exp}\left[\int^{\beta}_{0}d\tau\int{d^{3}x}\mathcal{L}_{\rm eff}\right]\times{\mathrm{det}\left(\frac{\partial{F}}{\partial{\omega}}\right)\delta{F}}, 
\end{eqnarray}
where the effective Lagrangian $\mathcal{L}_{\rm eff}$ of the theory is 
\begin{eqnarray}\label{effectlag}
\mathcal{L}_{\rm eff}&=&-\left[\partial_{\mu}-\frac{iq}{c\hbar}\left(A_{\mu}+\frac{\mu}{q}\eta_{\mu0}\right)\right]{\phi}^{*}\left[\partial^{\mu}+\frac{iq}{c\hbar}\left(A^{\mu}+\frac{\mu}{q}\eta^{\mu0}\right)\right]{\phi}-\frac{m^{2}c^{2}}{\hbar^{2}}\phi^{*}{\phi}-\lambda({\phi}^{*}{\phi})^{2}\nonumber\\
&&-\frac{1}{4}F_{\mu\nu}F^{\mu\nu}.
\end{eqnarray}

Before going on, a careful analysis in the effective Lagrangian (\ref{effectlag}) indicate that the shift
\begin{equation}\label{shift}
A_{\mu}\rightarrow{A_{\mu}}-\frac{\mu}{q}\eta_{\mu0}, 
\end{equation}
will provide an $\mu$-independent partition function. However, if the Lagrangian (\ref{effectlag}) does not have any $\mu-$dependence, the charged condensate will not be formed because the spontaneous symmetry breaking cannot occur since we are assuming $m^{2}>0$. To circumvent this problem, we can add to the effective Lagrangian a term like $q\mathcal{J_{\mu}}A^{\mu}$, where $\mathcal{J^{\mu}}(\equiv{J_{0}\eta^{\mu0}})$ is a constant background charge density. Physically, it means that this extra term will compensate the charge density of the scalar field, making the system electrically neutral, so thermodynamic equilibrium can be achieved and the charged condensate can emerge (For a further discussion, see \cite{Kapusta:1981aa}).

It follows that the effective Lagrangian $\mathcal{L}_{\rm eff}$ can be written as 
\begin{eqnarray}
\mathcal{L}_{\rm eff}&=&-\left(D_{\mu}{\phi}\right)^{*}\left(D^{\mu}
{\phi}\right)+\frac{i\mu}{\hbar{c}}\left({\phi}\partial_{0}
{\phi}^{*}-
{\phi}^{*}\partial_{0}
{\phi}\right)+2\frac{q\mu}{\hbar^{2}{c^{2}}}{A^{0}}
{\phi}^{*}
{\phi}-V(
{\phi})\nonumber\\\label{effectivelagrangian}
&&-\frac{1}{4}F_{\mu\nu}F^{\mu\nu}-\frac{q}{\hbar{c}}\mathcal{J_{\mu}}A^{\mu}, 
\end{eqnarray}
where 
\begin{eqnarray}\label{effectivepotential}
V(
{\phi})=-\left[\frac{\mu^{2}}{\hbar^{2}{c}^{2}}-\frac{m^{2}c^{2}}{\hbar^{2}}\right]\phi^{*}
{\phi}+\lambda(
{\phi}^{*}
{\phi})^{2} 
\end{eqnarray}
is the effective potential. It is clear from the effective potential $V({\phi})$ that the charged relativistic gas will form a condensate only when $\mu^{2}/(\hbar{c})^{2}>m^{2}c^{2}/\hbar^{2}$. We would like to remark that besides the many similarities between Bose--Einstein condensation and spontaneous symmetry breaking mechanism, the physical and mathematical details concerning these phenomena are not completely equivalent. For instance, if we assume $m^{2}<0$ in the Lagrangian density (\ref{lagrangiandensity}), the vacuum undergoes spontaneous symmetry breaking and the system ends up with a real scalar field and a massive real vector field. This is the usual Higgs mechanism. However, in the Higgs case, we cannot construct an acoustic geometric description for the perturbations since the conserved current for the real scalar field is identically null and the continuity equation is trivial. Nevertheless, as explained above, when we introduce the chemical potential $\mu$, the possibility of condensation exists \cite{Kapusta:1981aa}.

Now, one can apply the mean-field prescription by performing the substitution ${\phi}\rightarrow\varphi$, where $\varphi$ is the classical condensate field. 
So, performing the variation of $\varphi^{*}$ in the effective Lagrangian (\ref{effectivelagrangian}), the equation of motion for $\varphi$ assumes the form
\begin{equation}
\left[\Box-\frac{m^{2}c^{2}}{\hbar^{2}}+2\frac{i\mu}{\hbar{c}}\partial_{0}-\frac{\mu^{2}}{\hbar^{2}c^{2}}+2\frac{iq}{c\hbar}A^{\mu}\partial_{\mu}+\frac{iq}{c\hbar}\partial_{\mu}A^{\mu}-\frac{q^{2}}{c^{2}\hbar^{2}}A^{\mu}A_{\mu}-2\frac{q\mu}{\hbar^{2}c^{2}}{A_{0}}-U'(\varrho;\lambda)\right]\varphi=0,
\end{equation}
where $\varrho=\varphi^{*}\varphi$ and $U\equiv\lambda({\varphi}^{*}{\varphi})^{2}$ is the self-interaction term. Making a shift $A_{\mu}\rightarrow{A}_{\mu}-\frac{\mu}{q}\eta_{\mu0}$, we can factor out the chemical potential dependence and express the field equation as
\begin{equation}\label{fieldequationcrbec}
\left[\Box-\frac{m^{2}c^{2}}{\hbar^{2}}+2\frac{iq}{c\hbar}A^{\mu}\partial_{\mu}+\frac{iq}{c\hbar}\partial_{\mu}A^{\mu}-\frac{q^{2}}{c^{2}\hbar^{2}}A^{\mu}A_{\mu}-U'(\varrho;\lambda)\right]\varphi=0,
\end{equation}
which describes a charged relativistic Bose--Einstein condensate. 


\section{Dynamics of Perturbations}

We are interested in analyzing the dynamics of the linearized perturbation of the condensate. In a dilute gas, i.e., when the correlations in the gas can be neglected, we can use the mean-field approximation to compute the perturbations. So, applying the decomposition ${\phi}=\varphi(1+{\psi})$ in the modified Klein--Gordon equation (\ref{fieldequationcrbec}), where the classical condensate field $\varphi$ satisfies equation (\ref{fieldequationcrbec}) and ${\psi}$ is the relative quantum (i.e. of order $\hbar$) field fluctuation, we show that the linearized perturbations obey, at linear order of ${\psi}$,

\begin{equation}
\Box{\psi}+2\eta^{\mu\nu}(\partial_{\mu}ln\varphi)\partial_{\nu}{\psi}+2\frac{iq}{c\hbar}{A}^{\mu}\partial_{\mu}{\psi}-\varrho{U''}(\varrho;\lambda_{i})({\psi}+{\psi}^{*})=0. 
\end{equation}

It is very convenient to decompose the degrees of freedom of the complex mean-field $\varphi$ through the Madelung representation, given by $\varphi=\sqrt{\varrho}e^{i\theta}$, where $\varrho$ is the density and $\theta$ is the phase of the condensate. With that, we can define:
\begin{eqnarray}
u^{\mu}&\equiv&\frac{\hbar}{m}\eta^{\mu\nu}\partial_{\nu}\theta,\\
c_{0}^{2}&\equiv&\frac{\hbar^{2}}{2m^{2}}\varrho{U''}(\varrho;\lambda_{i}),\\
T_{\varrho}&\equiv&-\frac{\hbar^{2}}{2m}(\Box+\eta^{\mu\nu}\partial_{\mu}ln\varrho\partial_{\nu})=-\frac{\hbar^{2}}{2m\varrho}\eta^{\mu\nu}\partial_{\mu}\varrho\partial_{\nu}, 
\end{eqnarray}
where $c_{0}$ encodes the strength of the interactions and has dimensions of velocity and $T_{\varrho}$ is a generalized kinetic operator that in the non-relativistic limit and for constant $\rho$ reduces to
\begin{equation}
T_{\varrho}\rightarrow-\frac{\hbar^{2}}{2m\varrho}\nabla\varrho\nabla=-\frac{\hbar^{2}}{2m}\nabla^{2}, 
\end{equation}
which is the usual kinetic operator \cite{Garay:1999sk,Garay:2000jj,Mayoral:2010ck}.

In the phase-density decomposition, the locally conserved current and the condensate classical wave assume the form
\begin{eqnarray}\label{ce}
\partial_{\mu}\left(\varrho{f^{\mu}}\right)&=&0,\\
\label{classicalwavepdd}
-f^{\mu}f_{\mu}&=&c^{2}+\frac{\hbar^{2}}{m^{2}}\left[U'-\frac{\Box\sqrt{\varrho}}{\sqrt{\varrho}}\right],
\end{eqnarray}
where $f^{\mu}\left(\equiv{u}^{\mu}+\frac{q}{mc}{A}^{\mu}\right)$ is the timelike gauge invariant four-velocity of the condensate.

Using the above definitions, it is easy to show, following the steps of \cite{Fagnocchi:2010sn}, that the equation describing the propagation of charged perturbations is 
\begin{eqnarray}
\left[i\hbar{f^{\mu}}\partial_{\mu}-T_{\varrho}-mc_{0}^{2}\right]{\psi}=mc_{0}^{2}{\psi}^{*}. 
\end{eqnarray}

It is instructive to obtain a single equation for the quantum field ${\psi}$. This can be done taking the Hermitian conjugate of the above equation and using the result to eliminate ${\psi}^{*}$. After some manipulation, we find
\begin{eqnarray}\label{lpe}
\left\lbrace\left[i\hbar{f^{\mu}}\partial_{\mu}+T_{\varrho}\right]\frac{1}{c_{0}^{2}}\left[-i\hbar{f^{\nu}}\partial_{\nu}+T_{\varrho}\right]-\frac{\hbar^{2}}{\varrho}\eta^{\mu\nu}\partial_{\mu}\varrho\partial_{\nu}\right\rbrace{\psi}=0. 
\end{eqnarray}

The equation (\ref{lpe}) is a relativistic equation of motion describing the propagation of charged linearized perturbations on top of a charged rBEC. We call attention to the fact that the relative fluctuation ${\psi}$ does not change under a gauge transformation, which assures us that equation (\ref{lpe}) is indeed gauge invariant as expected. We also remark that the electromagnetic field under consideration is endowed with a small magnitude. This assumption is very convenient since we intend to neglect background effects coming from electromagnetic fluctuations in the phonon propagation.


\section{The Relativistic Acoustic Metric}

One of the usual assumptions in analogue gravity is the flow be locally irrotational, i.e., vorticity free. Nevertheless, when the condensate couples with the electromagnetic field, the situation changes and vorticity appears in the system. It can be seen explicity by inspection of the vorticity tensor $w_{\mu\nu}$ contracted with the charged four-velocity $f^{\mu}$. Introducing the projection tensor $h^{\mu}_{\nu}=\delta^{\mu}_{\nu}+f^{\mu}f_{\nu}/f^{2}$, where $f^{2}=f^{\alpha}f_{\alpha}$, this is given by 
\begin{eqnarray}
w_{\mu\nu}=h^{\alpha}{}_{\mu}h^{\beta}{}_{\nu}\nabla_{[\alpha}f_{\beta]}=h^{\alpha}{}_{\mu}h^{\beta}{}_{\nu}\frac{q}{mc}\left(\partial_{\alpha}A_{\beta}-\partial_{\beta}A_{\alpha}\right).
\end{eqnarray}
One can easily see that the vorticity tensor $w_{\mu\nu}$ cannot be set generically equal zero since the field strength tensor $F_{\mu\nu}$ is generically non zero and also the components of its projection on the spacelike hypersurface orthogonal to $f^{\mu}$.

The question that arises is the following: Is it still possible describe charged linearized perturbations in the formalism of curved spacetime in the presence of vorticity? Remarkably, the answer is yes. Let us specify carefully the conditions under which this can be achieved.

First of all, we note that the charged rBECs lead to a excitation spectra similar to the uncharged case analyzed by \cite{Fagnocchi:2010sn}. The difference lies in the fact that the electromagnetic minimal coupling modifies the uncoupled conserved current due the presence of the gauge field, which introduces a shift in the four-velocity $u^{\mu}$, namely,
\begin{equation}
j^{\mu}=\rho\frac{m}{\hbar}u^{\mu}\rightarrow\rho\frac{m}{\hbar}\left(u^{\mu}+\frac{q}{mc}A^{\mu}\right).
\end{equation}
Despite having a complicated excitation spectra,  for sufficiently low momenta the phonon propagation described in equation (22) leads to two quasiparticle modes, a massive and a massless, as in uncoupled case. Bearing in mind that we want to find an energy regime where we can apply the analogue gravity framework to describe phonon propagation in the presence of vorticity, we will focus on the gapless modes. The appropriate limit which the aforementioned framework can be acquired needs to satisfy basically two conditions: 
\begin{enumerate}[(i)]
 \item one needs to be within the so called the phononic regime, the low momenta range for the gapless excitation, i.e. \cite{Fagnocchi:2010sn}
\begin{equation}\label{lowmomenta}
|\mathbf{k}|\ll2\frac{mc_{0}}{\hbar}\left[1+\left(\frac{c_{0}}{f^{0}}\right)^{2}\right], 
\end{equation}
where the term on the right is a relativistic generalization of the inverse of the healing length for charged condensates, and 
\item one should be able to neglect the quantum potential $T_{\varrho}$ in equation (\ref{lpe}), which can be achieved assuming that the background quantities varies slowly in space and time on scales comparable with the wavelength $w$ of the perturbations, conditions that can be written as
\begin{equation}\label{neglectingqp}
\left|\frac{\partial_{t}\rho}{\rho}\right|\ll{w}, \quad\quad\quad \left|\frac{\partial_{t}c_{0}}{c_{0}}\right|\ll{w}, \quad\quad\quad \left|\frac{\partial_{t}f_{\mu}}{f_{\mu}}\right|\ll{w}.  
\end{equation}
\end{enumerate}

The previous considerations reduce equation (\ref{lpe}) to
\begin{eqnarray}\label{pel}
&&\left[{f}^{\mu}\partial_{\mu}\frac{1}{c_{0}^{2}}{f^{\nu}}\partial_{\nu}-\frac{1}{\varrho}\eta^{\mu\nu}\partial_{\mu}\varrho\partial_{\nu}\right]{\psi}=0.
\end{eqnarray}

Multiplying (\ref{pel}) by $\varrho$ and using the continuity equation (\ref{ce}), we have
\begin{eqnarray}\label{linearizedeqnoqp}
\partial_{\mu}\left[\frac{\varrho}{c_{0}^{2}}{f^{\mu}}{f^{\nu}}-\varrho\eta^{\mu\nu}\right]\partial_{\nu}{\psi}=0. 
\end{eqnarray}

It is obvious that one can express the above equation (\ref{linearizedeqnoqp}) as
\begin{equation}
\partial_{\mu}\left(\gamma^{\mu\nu}\partial_{\nu}{\psi}\right)=0, 
\end{equation}
where $\gamma^{\mu\nu}$ is 
\begin{center}
$\gamma^{\mu\nu}=\frac{\rho}{c_{0}^{2}}\left[\begin{array}{cc}
-c_{0}^{2}-(f^{0})^{2} & -f^{0}f^{j}\\
-f^{0}f^{i} & c_{0}^{2}\delta^{ij}-f^{i}f^{j}
\end{array}\right].$
\end{center}

If one identifies $\gamma^{\mu\nu}=\sqrt{-g}g^{\mu\nu}$, then
\begin{equation}
\sqrt{-g}=\rho^{2}\sqrt{1-f^{\alpha}f_{\alpha}/c_{0}^{2}}, \quad 
\end{equation}
and
\begin{center}
$g^{\mu\nu}=\frac{1}{\rho{c}_{0}^{2}\sqrt{1-f^{\alpha}f_{\alpha}/c_{0}^{2}}}\left[\begin{array}{cc}
-c_{0}^{2}-(f^{0})^{2} & -f^{0}f^{j}\\
-f^{0}f^{i} & c_{0}^{2}\delta^{ij}-f^{i}f^{j}
\end{array}\right].$
\end{center}

Therefore, equation (\ref{linearizedeqnoqp}) can be cast in the form  
\begin{equation}\label{dalembertian}
\triangle{\psi}\equiv\frac{1}{\sqrt{-g}}\partial_{\mu}\left(\sqrt{-g}g^{\mu\nu}\partial_{\nu}{\psi}\right),
\end{equation}
which is a d'Alembertian in a curved background. From the above equation, we promptly realize that the quasiparticle propagation in a nonhomogeneous fluid can be described by a relativistic equation of motion in a curved acoustic spacetime, where the emergent geometry is determined by the acoustic metric $g_{\mu\nu}$. Inverting $g^{\mu\nu}$, one 
can then see that the acoustic metric $g_{\mu\nu}$ for phonons propagation in a (3+1)D relativistic, barotropic, {\it rotational} fluid flow is given by
\begin{eqnarray}\label{relativisticacousticmetric}
g_{\mu\nu}=\frac{\varrho}{\sqrt{1-f_{\alpha}f^{\alpha}/c_{0}^{2}}}\left[\eta_{\mu\nu}\left(1-\frac{f_{\alpha}f^{\alpha}}{c_{0}^{2}}\right)+\frac{f_{\mu}f_{\nu}}{c_{0}^{2}}\right]. 
\end{eqnarray}

We remark that the acoustic metric (\ref{relativisticacousticmetric}) is gauge invariant. Another useful way to express the relativistic acoustic metric (\ref{relativisticacousticmetric}) is using the definitions
\begin{eqnarray}
v^{\mu}&=&c\frac{f^{\mu}}{||f||}, \quad\quad ||f||=\sqrt{-\eta_{\mu\nu}f^{\mu}f^{\nu}}, 
\end{eqnarray}
where $v^{\mu}$ is the normalized four-velocity and $||f||$ is the normalization factor. With that, the relativistic acoustic metric in the presence of vorticity assumes the form
\begin{eqnarray}\label{relacmetr}
g_{\mu\nu}=\frac{\varrho{c}}{c_{s}}\left[\eta_{\mu\nu}+\left(1-\frac{c_{s}^{2}}{c^{2}}\right)\frac{v_{\mu}v_{\nu}}{c^{2}}\right], 
\end{eqnarray}
which is disformally related to the background Minkowski spacetime and the speed of sound $c_{s}$ is defined as
\begin{eqnarray}\label{speedofsound}
c_{s}^{2}=\frac{c^{2}c_{0}^{2}/||f||^{2}}{1+c_{0}^{2}/||f||^{2}}. 
\end{eqnarray}

The relativistic acoustic metric (\ref{relacmetr}) describes in a simple fashion way perturbations in a charged rBEC which include vorticity. It is important to mention that this is not the first attempt to incorporate vorticity in the context of analogue gravity. An approach through the use of Clebsch potentials can be found in \cite{PerezBergliaffa:2001nd}. Nevertheless, the wave equation generated is more complicated than a simple d'Alembertian and the construction significantly more difficult. We further stress that at the level of geometrical acoustics one may incorporate viscosity. However, geometrical acoustics is not enough for many purposes, and an irrotational fluid is a crucial assumption in deriving the wave equation on the effective curved metric.

\section{The non-relativistic charged Bose--Einstein condensate}
 
We have shown in the previous section that linearized perturbations over a relativistic charged condensate can, under suitable assumptions, be described in the same way as a scalar field propagating in a curved spacetime. Now, we can therefore ask if this description can be achieved when the charged condensate is non-relativistic. To see if such a description can indeed be done, we will start considering the non-relativistic limit of the relativistic equation of motion (\ref{lpe}).

To begin with, in the non-relativistic regime, the external interaction $qA^{0}$ is much smaller then the atomic's rest energy $mc^{2}$, namely $qA^{0}\ll{m}c^{2}$. Moreover, the self-interaction between the atoms in the condensate must be weak, which means that $c_{0}\ll{c}$. It is also easy to see from equation (\ref{classicalwavepdd}) that in the non-relativistic regime $f^{0}\rightarrow{c}$. In addition, the speed of sound $c_{s}$ defined by relation (\ref{speedofsound}) reduces to $c_{0}$. 

Assuming that these conditions are satisfied, the condition (\ref{lowmomenta}) is in turn given by 
\begin{equation}\label{nonrelativisticphononic}
|\mathbf{k}|\ll2\frac{mc_{0}}{\hbar},
\end{equation}
which determine the momenta scale where the acoustic description can be applied. 

Taking into account the previous considerations in the equation of motion (\ref{lpe}) we arrive at
\begin{eqnarray}\label{nrpe}
\left\lbrace\left[{i}\hbar\left(\partial_{t}+f^{i}\partial_{i}\right)+T_{NR}\right]\frac{1}{c_{0}^{2}}\left[-{i}\hbar\left(\partial_{t}+f^{j}\partial_{j}\right)+T_{NR}\right]-\frac{\hbar^{2}}{\rho}\nabla\rho\nabla\right\rbrace{\psi}=0,
\end{eqnarray}
where $f^{i}(\equiv{v^{i}}+\frac{q}{mc}A^{i})$ is the charged 3-velocity the condensate, $v^{i}$ is the velocity of the condensate in the uncharged case and the standard quantum potential in the non-relativistic limit $T_{NR}$ is
\begin{equation}
T_{NR}\equiv-\frac{\hbar^{2}}{2m\rho}\nabla\rho\nabla,
\end{equation}
where $\rho$ is the mass density.

The phonon propagation under an effective acoustic metric can be done only when one can neglect the quantum potential $T_{NR}$, which can be achieved under the assumptions (\ref{neglectingqp}). In this case, the equation (\ref{nrpe}) reduces to 
\begin{eqnarray}\label{reducednrpe}
\left[\left(\partial_{t}+f^{i}\partial_{i}\right)\frac{1}{c_{0}^{2}}\left(\partial_{t}+f^{j}\partial_{j}\right)-\frac{1}{\rho}\nabla\rho\nabla\right]{\psi}=0.
\end{eqnarray}

Multiplying (\ref{reducednrpe}) by the mass density $\rho$ and using the continuity equation
\begin{equation}
\partial_{t}\rho+\partial_{i}\left(\rho{f^{i}}\right)=0,
\end{equation}
we promptly arrive at
\begin{eqnarray}
-\partial_{t}\left[\frac{\rho}{c_{0}^{2}}\left(\partial_{t}{\psi}+f^{j}\partial_{j}{\psi}\right)\right]+\partial_{i}\left[\rho\partial^{i}{\psi}-\frac{\rho}{c_{0}^{2}}f^{i}\left(\partial_{t}{\psi}+f^{j}\partial_{j}{\psi}\right)\right]=0,
\end{eqnarray}
which has the same structure that the wave equation describing the propagation of acoustic disturbances in irrotational fluids [See section 2.3 of \cite{Barcelo:2005fc}]. Following the standard methodology of analogue gravity, we find the usual non-relativistic acoustic metric
\begin{center}
\begin{equation}\label{nonrelativisticacousticmetric}
g_{\mu\nu}=\frac{\rho}{c_{s}}\left[\begin{array}{cc}
-(c_{s}^{2}-\mathbf{f}^{2}) & -f_{j}\\
-f_{i} & \delta_{ij}
\end{array}\right],
\end{equation}
\end{center}
where $\mathbf{f}^{2}=f_{i}f^{i}$ is the squared 3 velocity of the fluid flow.

In non-relativistic fluids the locally vorticity free condition over the velocity flow vector $\mathbf{v}$ is assured by the requirement $\mathbf{\nabla}\times{\mathbf{v}}=\mathbf{0}$. In our prescription, we have derived the non-relativistic acoustic metric without making use of the vorticity free assumption. To see explicity that the system is endowed with vorticity, we note that the rotational of the charged 3-velocity $\mathbf{f}$ is
\begin{equation}
\mathbf{\nabla}\times\mathbf{f}=\frac{q}{mc}\mathbf{\nabla}\times\mathbf{A}=\frac{q}{mc}\mathbf{B},
\end{equation}
which is clearly non zero and implies the existence of vorticity in the BEC. 

A point that deserves a careful attention is the absence of the $A^{0}$ component as a background quantity in the non-relativistic acoustic metric (\ref{nonrelativisticacousticmetric}). In the framework of hydrodinamical systems, external potentials do not influence the description of the linearized perturbations. As a physical consequence in the BEC, even when the charged condensate is under the action of a static electric field, the phonons propagation are insensitive to it. 

Now, in order to check that our derivation of the non-relativistic acoustic metric in the presence of vorticity is correct, one can take the alternative route to impose the electromagnetic minimal coupling directly to the equation that describes the condensate in the non-relativistic limit, namely,
\begin{equation}\label{nonrelativisticbec}
i\hbar\partial_{t}\phi=\frac{1}{2m}\left(-i\hbar\partial_{i}\right)^{2}\phi+V_{ext}+g|\phi|^{2}\phi, 
\end{equation}
where $V_{ext}$ is an external potential, $m$ is the mass of the bosons, $|\phi|^{2}$ is the atomic density and $g$ is the effective coupling constant which describes locally the scattering of atoms. The equation (\ref{nonrelativisticbec}) emerges quite naturally in the analysis of BEC up to a first order approach, and it is formally equivalent to the Schr{\"o}dinger equation with a nonlinear term $g|\phi|^{2}$. It is worth noting that the Ginzburg-Landau theory of superconductivity \cite{Ginzburg:1950sr} is a particular case of (\ref{nonrelativisticbec}). Performing the electromagnetic minimal coupling, the equation (\ref{nonrelativisticbec}) assumes the form
\begin{eqnarray}\label{cgpe}
i\hbar\partial_{t}\phi&=&\frac{1}{2m}\left(-i\hbar\partial_{i}+\frac{q}{c}A^{i}\right)^{2}\phi+qA^{0}\phi+g|\phi|^{2}\phi.
\end{eqnarray}

Proceeding exactly the same way as in the relativistic case in order to obtain the equation for ${\psi}$, we insert ${\phi}=\varphi(1+{\psi})$ in equation (\ref{cgpe}) and get, at the linearized level,
\begin{eqnarray}
&&\left[i\hbar\partial_{t}+\frac{\hbar^{2}}{2m}\nabla^{2}+\frac{\hbar^{2}}{m}(\partial_{i}ln\varphi)\partial_{i}+i\hbar\frac{{q}}{mc}A^{i}\partial_{i}-mc_{sNR}^{2}\right]{\psi}=mc_{sNR}^{2}{\psi}^{*},
\end{eqnarray}
where $c_{sNR}^{2}=g\rho/m$ is the speed of sound in the non-relativistic BEC.

Again, decomposing the condensate wave function as $\varphi=\sqrt{\rho}e^{i\theta}$, where $\rho$ is the mass density, and rewriting to obtain a single equation to ${\psi}$, we get
\begin{eqnarray}
\left\lbrace\left[{i}\hbar(\partial_{t}+f^{i}\partial_{i})+T_{NR}\right]\frac{1}{c_{sNR}^{2}}\left[-{i}\hbar(\partial_{t}+f^{j}\partial_{j})+T_{NR}\right]-\frac{\hbar^{2}}{\rho}\nabla\rho\nabla\right\rbrace{\psi}=0,
\end{eqnarray}
which is exactly the same as equation (\ref{nrpe}). Employing the same conditions that previously discussed and following the usual steps, we arrive at the non-relativistic acoustic metric (\ref{nonrelativisticacousticmetric}), so lending support to our previous derivation.

\section{Conclusions}

One standard limitation with the procedures of analogue gravity is that the derivation of the wave function for the acoustic disturbances is possible only when we have a barotropic and inviscid fluid and the flow is irrotational. Under these assumptions, one can derive a d'Alembertian equation of motion to describe the linearized perturbations, which is identical to a relativistic scalar massless field propagating in a curved spacetime. 

In this work we have shown that is possible to overcome the irrotational constraint in moving fluids and incorporate vorticity in the description of sound propogation in condensates. Performing the electromagnetic minimal coupling to the BEC we found that the conserved current related to gauge $U(1)$-symmetry depends on the complex scalar and gauge fields. Therefore the requirement that the gauge invariant charged flow velocity be locally irrotational is no longer imposed as the vorticity tensor is generically non-zero. So, taking into account the low momentum limit of the charged BEC and in the regime where one can neglect the quantum potential, the propogation of the charged quasiparticles in the presence of vorticity can be described by the formalism of the quantum field theory in a curved spacetime. Without these assumptions, the charged phonons propagation are described by a complicated differential wave equation. We emphasize that both cases (relativistic and non-relativistic) have gauge invariant equations for perturbations.

We also would like to remark that the condition on the momenta (\ref{nonrelativisticphononic}) which defines the phononic regime in the non-relativistic limit is exactly the same as required for the uncharged BEC \cite{Fagnocchi:2010sn}. As in the uncoupled case, the momenta scale in which the analogue framework can be applied depends on the inverse of the healing length $mc_{0}/\hbar$. This coincidence can be easily comprehended by noting that according to (\ref{lowmomenta}), which determines the phononic regime in the relativistic range, when one takes into account the non-relativistic limit, $u^{0}\approx{c}$ and $A^{0}$ becomes a negligible term, implying that $(c_{0}/f^{0})^2\rightarrow{c}_{0}^{2}/c^{2}\ll1$. In this case we see that the relativistic phononic regime (\ref{lowmomenta}) reduces to (\ref{nonrelativisticphononic}). As previously discussed in section V, in the non-relativistic limit, static electric fields contributes to the dynamics of the background equations of motion, but does not influence the quasiparticles' propagation. 

The cheif value of this work, with respect to the previous attempts to incorporate vorticity is the simplicity, both in the technical details of the construction, and in the physical picture of vorticity arising from the action of a magnetic field on a charged flow. As such, it is a clear demonstration that it is possible to easily incorporate at least some systems with vorticity into the analogue framework.

\acknowledgments

The authors are grateful to Matt Visser for illuminating discussions and useful comments on the manuscript. The authors are also grateful to S. Shankaranarayanan for comments on an earlier version of the manuscript. Bethan Cropp is supported by Max Planck-India Partner Group on Gravity and Cosmology. Rodrigo Turcati is very grateful to CNPq for financial support. This publication was made possible through the support of a grant from the John Templeton Foundation. The opinions expressed in this publication are those of the authors and do not necessarily reflect the views of the John Templeton Foundation.

\end{document}